\documentclass[fleqn,usenatbib]{mnras}

\usepackage{newtxtext,newtxmath}

\usepackage[T1]{fontenc}
\usepackage{ae,aecompl}

\newcommand{\ms}{\mbox{m\,s$^{-1}$}}
\newcommand{\gtsimeq}{\raisebox{-0.6ex}{$\,\stackrel
         {\raisebox{-.2ex}{$\textstyle >$}}{\sim}\,$}}
\newcommand{\ltsimeq}{\raisebox{-0.6ex}{$\,\stackrel
         {\raisebox{-.2ex}{$\textstyle <$}}{\sim}\,$}}

\usepackage{graphicx}	
\usepackage{amsmath}	
\usepackage{amssymb}	
\usepackage{subfigure}
\usepackage{xcolor}

\title[Truly eccentric II]{Truly eccentric. II. When can two circular planets mimic a single eccentric orbit?}

\author[R.A. Wittenmyer et al.]
{Robert A. Wittenmyer,$^{1}$\thanks{E-mail: rob.w@usq.edu.au (RW)}
Christoph Bergmann,$^{2}$
Jonathan Horner,$^{1}$
Jake Clark,$^{1}$
\newauthor Stephen R. Kane,$^{3}$
\\
$^{1}$University of Southern Queensland, Centre for Astrophysics, Toowoomba QLD 4350, Australia\\
$^{2}$Exoplanetary Science at UNSW, School of Physics, UNSW Sydney, NSW 2052, Australia\\
$^{3}$Department of Earth Sciences, University of California Riverside, 900 University Avenue, Riverside, CA 92521, USA
}

\date{Accepted XXX. Received YYY; in original form ZZZ}

\pubyear{2018}

\begin{document}
\label{firstpage}
\pagerange{\pageref{firstpage}--\pageref{lastpage}}
\maketitle

\begin{abstract}
When, in the course of searching for exoplanets, sparse sampling and noisy data make it necessary to disentangle possible solutions to the observations, one must consider the possibility that what appears to be a single eccentric Keplerian signal may in reality be attributed to two planets in near-circular orbits. There is precedent in the literature for such outcomes, whereby further data or new analysis techniques reveal hitherto occulted signals.  Here, we perform suites of simulations to explore the range of possible two-planet configurations that can result in such confusion. We find that a single Keplerian orbit with $e\gtsimeq$0.5 can virtually never be mimicked by such deceptive system architectures. This result adds credibility to the most eccentric planets that have been found to date, and suggests that it could well be worth revisiting the catalogue of moderately eccentric 'confirmed' exoplanets in the coming years, as more data become available, to determine whether any such deceptive couplets are hidden in the observational data.

\end{abstract}




\section{Introduction}

Roughly thirty years ago, we saw the dawn of the Exoplanet Era, with the detection of the first planet-mass objects orbiting other stars \citep{GammaCeph,Latham,psr1257,51peg}. In the decades since, we have become ever more adept at observing the minute variations in the behaviour of stars that hint at the presence of their planetary companions \citep{fischer16}.

Through the late 1990s and early 2000s, the radial velocity (RV) technique dominated exoplanetary science, revealing a variety of planets that was far greater than we had previously imagined. We found giant, Jupiter-mass planets orbiting perilously close to their host stars \citep[e.g.][]{HJ1,HJ2,HJ3}, as well as planets moving on orbits so eccentric that they more closely resemble the orbits of comets than the planets in our own backyard \citep[e.g.][]{ecc1, ecc2, 76920}. We also found many systems with multiple planets moving on orbits locked in mutual mean-motion resonance \citep[e.g.][]{res2,res3,res1,res4,res5}. 

Over the past decade, it has become clear that it is possible for a multiple-planet system containing resonant planets on near-circular orbits to masquerade as a single, moderately eccentric planet in typical sparsely sampled radial velocity data \citep[e.g.][]{ang10, songhu, boisvert18}.  Since researchers typically look for the simplest explanation for a given signal, such systems are often initially reported as single, moderately eccentric planets.  The true multiplicity of these systems is then only revealed after further observations are carried out \citep[e.g.][]{witt12,kurster15, trifonov17}.  For this reason, in 2013, we carried out a pilot study examining the likelihood that several moderately eccentric exoplanetary systems within the published literature might actually be such multiple planet systems, masquerading as single worlds \citep{songhu}.  On the other hand, a number of extremely eccentric planets have been found \citep[e.g.][]{ecc1,ecc2,ecc4,ecc5,76920}, whose best-fit orbits are so extreme that it seems unlikely that they could be reproduced by any given combination of two exoplanets moving on resonant, near circular orbits.

This, then, poses an obvious question - how eccentric can an orbit be before one can be truly confident that we are observing a single planet on a highly eccentric orbit, rather than poorly sampling a multiple planet system. Clearly, there exists a threshold for which no multiple planet solution can explain a highly eccentric orbit. Equally, there exists a range of single planet orbital eccentricities that could readily be explained by invoking multiple planets on near circular orbits.

In this work, we attempt to answer that question, in order to both strengthen confidence in the interpretation of extremely eccentric single planet systems, and to identify the most dangerous regime of eccentricity space, for which the risk is greatest that a given multiple planet system will be misidentified as a single, moderately eccentric world.

In Section~\ref{Approach}, we describe the approach we take to address this question, detailing how we created simulated radial velocity data sets to model the effects of multiplicity and sparse data sampling on the type of solutions found for a given planetary system. In Section~\ref{Results}, we present the results of our analysis, before moving on to discuss those results, and draw our conclusions, in Section~\ref{Conclusions}.

\section{Simulation Approach}
\label{Approach}

In this section, we describe the procedures for producing the simulated 
radial velocity data sets. The stochasticity of real observational 
sampling in combination with the well-known sampling biases induced by 
telescope scheduling constraints \citep{monster} can result in poor 
detectability at certain orbital periods.  Perversely, we wish to 
embrace this real-life pathology to provide the best possible assessment of the 
degree to which observers might be bamboozled by circular double systems 
masquerading as single eccentric planets.



\subsection{Sampling}

To create simulated observation times for this experiment, we attempted 
to reproduce the sampling properties of real radial velocity data.  
Following the simulation procedure in \citet{witt13}, we made the 
following assumptions: (1) one observation in a 10-night block every 30 
days (bright-time scheduling), (2) the target is unobservable for four 
consecutive months every year, and (3) poor weather randomly prevents 
the observation 33\% of the time.  These conditions were selected 
firstly because planet-search programs are usually allocated time in 
bright lunations owing to the brightness of the targets, and secondly, 
for a mid-latitude site such as the Anglo-Australian Telescope, with 
planet-survey targets distributed randomly in Right Ascension, the 
average target is unobservable for four months in a year\footnote{Here 
we define ``unobservable'' to mean that the target spends less than one 
hour at an airmass less than 2.}.  This procedure generated a string of 
50 observation epochs for each of 10,000 simulated stars, for each 
scenario tested herein.

We also explored more realistic sampling by drawing the observation 
times from real data sets.  The 18-year Anglo-Australian Planet Search \citep[AAPS; e.g.][]{carter03,tinney11,newjupiters,30177} has 90 stars for which more than 50 epochs were obtained.  We 
generated strings of 50 epochs as follows: for each of the 90 real AAPS 
data files, we selected a 50-epoch window, then frame-shifted it by one 
until reaching the end of the data set.  In this way, a real data set 
with, e.g. 55 epochs would generate 6 lists of 50-epoch samples, 
preserving the sampling characteristics of the real data.  The result is 
3871 lists of 50 observational epochs, each drawn from a real AAPS 
target and hence preserving all the associated idiosyncrasies of real 
data.  Sets of 50 observation times for each of the 10,000 simulated data sets were 
then drawn at random (with replacement) from this pool.  Whilst the replacement means that some simulated datasets had identical observation times, we emphasize that the simulated velocity measurements are different, as described below. 

\subsection{Noise Model}

We simulated stellar velocity noise by choosing the radial velocity 
values and their uncertainties by a random draw from the AAPS data for 
six stable solar-type stars (531 epochs).  These velocities have a mean 
of zero and an rms scatter of 2.99\,\ms.  In this way, we assume the 
input data are purely noise containing no planetary signals, and we have 
made no assumptions about the noise distribution (e.g.~Gaussianity).  
The uncertainties, derived only from photon statistics, have a mean of 1.1\,\ms.  We then add 3\,\ms\ of stellar jitter in quadrature to the individual measurement uncertainties.  From our experience in least-squares fitting, this treatment reduces the possibility of the fitting routine getting stuck in local minima due to high-leverage points with small uncertainties.  In the next subsection, we describe the process for generating the simulated Keplerian signals, which are added to the noise to produce the final simulated data.

\subsection{Simulated Planetary Signals}

For all simulated two-planet systems, the velocity amplitude $K$ for 
each planet was assigned a random value between 20-100\,\ms.  This is 
largely arbitrary, but reflects values typical of securely detected 
radial velocity exoplanets.  That is, the amplitudes are large enough to 
be unambiguously detectable in the presence of noise, yet small enough 
to remain in the planetary regime.  Each scenario resulted in 10,000 
synthetic radial velocity data sets.

\textit{Scenario I} -- First we considered a simple circular-double 
configuration, with orbits whose periods are in a 2:1 commensurability (Scenario Ia) at 
arbitrarily chosen periods of 100 and 200 days.  In light of the 
observed pile up of planets at the 2.17:1 period ratio 
\citep{steffen15}, we repeated this with planets at periods of 217 and 
100 days, respectively (Scenario Ib).  

\textit{Scenario II} -- Next, we considered the combination of planets moving on slightly 
eccentric orbits.  Scenario IIa consists of two planets, both fixed at 
$e=0.1$, and at periods of 100 and 200 days as above.  Likewise, 
Scenario IIb considers planets on periods of 217 and 100 days.  For these eccentric orbits, we set the periastron arguments at $\omega_1=0$ and $\omega_2$ as a random value on $[0,2\pi]$.  Hence, the relative apsidal alignment of the two planets is randomised.

\textit{Scenario III} -- Finally, we investigated a more realistic 
choice of orbital periods for two circular planets.  In Scenario IIIa, 
we drew the period of the outer planet at random (with replacement) from 
a set of 673 planets detected by radial velocity, obtained from the NASA 
Exoplanet Archive.  The inner planet signal was then chosen to be 
exactly half this period.  As above, Scenario IIIb sets the inner planet 
period such that the two are in a 2.17:1 period ratio.

For each of the three scenarios above, we also created 'c' and 'd' subsets.  The 'c' and 'd' subsets of the above scenarios (I, II and III) are analogous to the 'a' and 'b' setups (2:1 and 2.17:1 respectively), but using the more realistic observation epoch times drawn from real data as described in Section 2.1.

\subsection{Orbit Fitting}

Armed with synthetic data sets, we proceeded to fit each with a single Keplerian signal using the IDL package \texttt{RVLIN} \citep{rvlin}, which employs the Levenberg-Marquardt method for non-linear $\chi^2$-minimisation. We obtained uncertainty estimates for the orbital parameters with the bootstrapping algorithm from the \texttt{BOOTTRAN} package \citep{wan12}.  As pointed out by \citet{wan12}, we note that bootstrapping is not an ideal method to determine parameter uncertainties in cases of sparsely sampled data, but our philosophy was to perform the fitting as blindly as possible. In order to save computational time, we limited our calculations to 1000 bootstrap realisations; comparison of a small number of data sets for which we also obtained uncertainty estimates using 100\,000 steps showed only marginal differences. The initial guess for the orbital period comes from the highest peak in a Lomb-Scargle periodogram \citep{lomb, scargle, hoba86}, and we used a default initial value of 0.3 for the eccentricity. We also employed an upper limit of 10\,000 days for the orbital period. The only deviation from a completely blind fit we pursued was to prevent the fitting routine from getting stuck at $e=0$, which sometimes happened especially if the fit was quite poor, and which we interpret as a peculiarity of the specific fitting package used.  While not excluding circular orbits entirely, we automatically stepped through initial guesses of the periastron passage time if the initial fit returned an eccentricity of zero.

As a sanity check, we also attempted to fit each synthetic data set with a two-planet model, keeping the eccentricities fixed at either zero or 0.1, depending on the scenarios described above.  Table~\ref{tab:sanitycheck} shows the degree to which this test successfully recovered the input periods of both planets to within 10\%.

\begin{table}
        \centering
        \caption{Recovery of two-planet solutions}
        \label{tab:sanitycheck}
        \begin{tabular}{lr}
                \hline
                Scenario & Fraction Recovered \\
                \hline
                Ia & 100.00\% \\
                Ib & 100.00\% \\
                Ic & 99.93\% \\
                Id & 99.96\% \\
                IIa & 100.00\% \\
                IIb & 100.00\% \\
                IIc & 99.98\%  \\
                IId & 99.90\%  \\
                IIIa & 98.69\% \\
                IIIb & 98.94\% \\
                IIIc & 91.73\% \\
                IIId & 92.94\% \\
                \hline
        \end{tabular}
\end{table}


\section{Results}
\label{Results}

\begin{figure*}
\includegraphics[width=\linewidth]{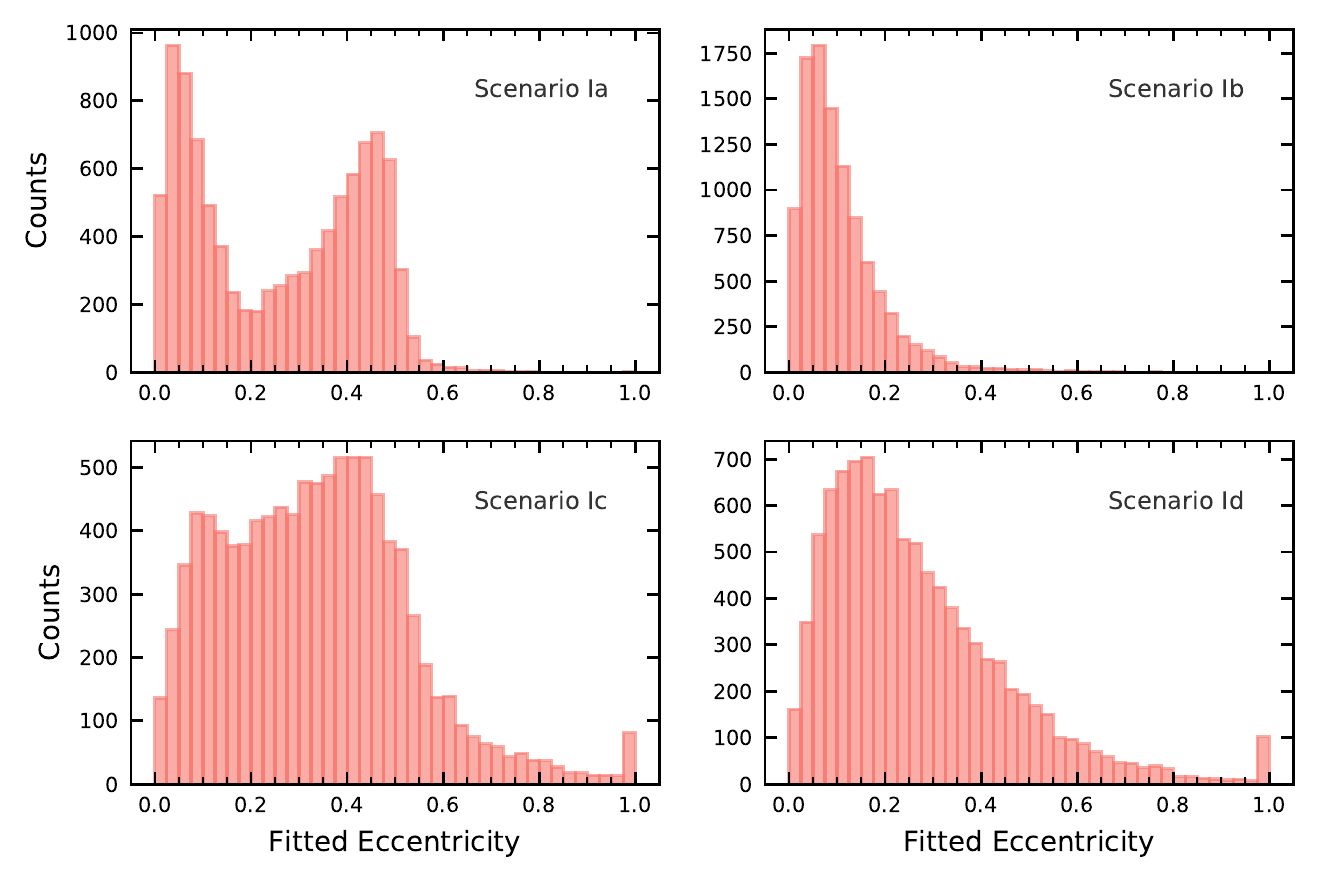}
\caption{Distribution of fitted eccentricities for Scenario I (Two circular planets). }
\label{fig:eccs1}
\end{figure*}

\begin{figure*}
\includegraphics[width=\linewidth]{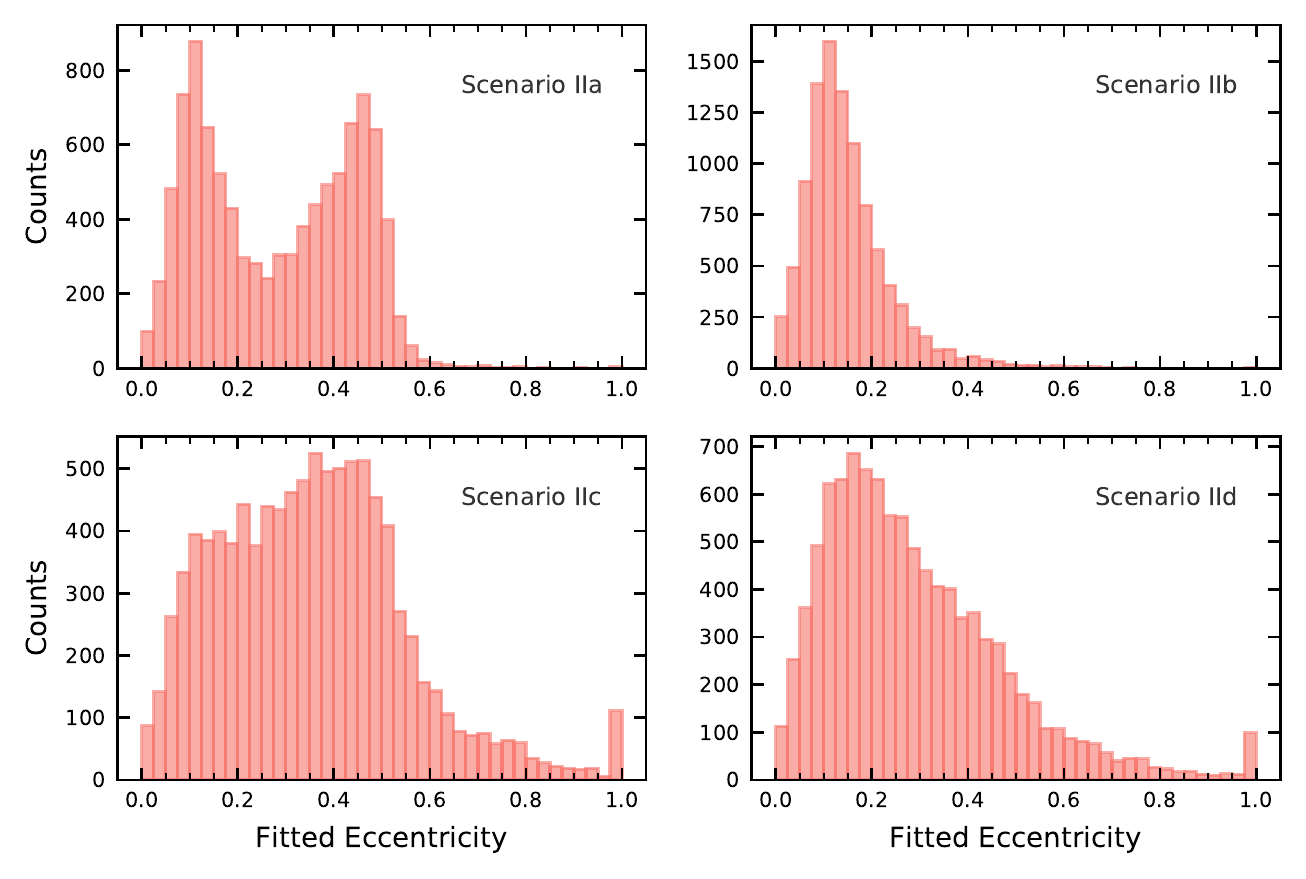}
\caption{Distribution of fitted eccentricities for Scenario II (Two slightly eccentric planets). }
\label{fig:eccs2}
\end{figure*}

\begin{figure*}
\includegraphics[width=\linewidth]{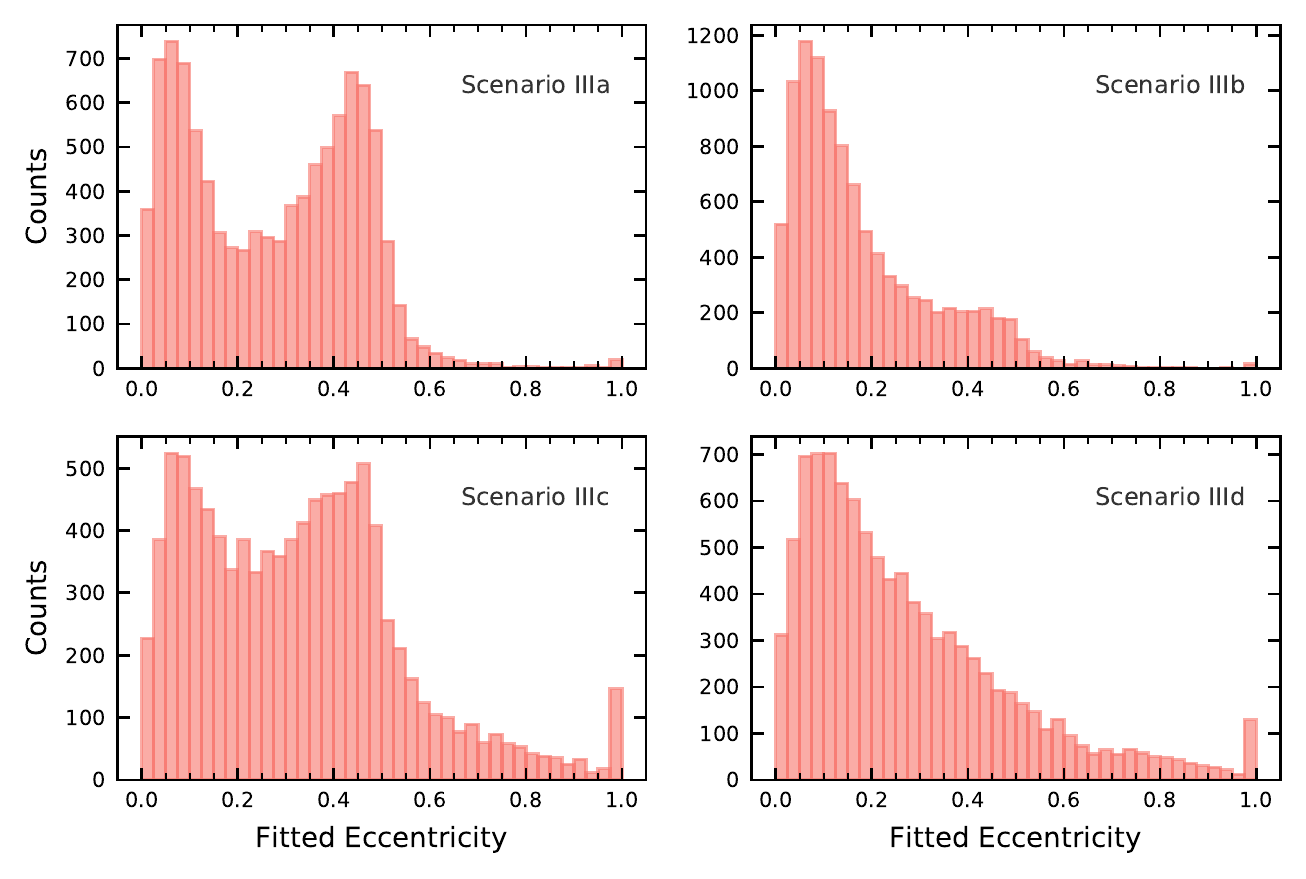}
\caption{Distribution of fitted eccentricities for Scenario III (Two circular planets with 'realistic' orbital periods). }
\label{fig:eccs3}
\end{figure*}

Not surprisingly, the act of fitting a single planet to data containing 
two signals produced a wide range of results.  In the present work, we 
are most concerned with the eccentricity; 
Figures~\ref{fig:eccs1}--\ref{fig:eccs3} show the distributions of 
fitted eccentricities resulting from all scenarios.

\clearpage

\clearpage

\clearpage

The primary aim of this work is to determine the frequency and 
conditions which cause two circular (or nearly-circular) planets to 
masquerade as a single eccentric planet in realistically-sampled radial 
velocity data.  To quantify this, we must first set some criteria for a 
``plausible'' single-eccentric fit.  As the input data consist of two 
Keplerian signals at two different periods, which we fit with a single 
planet, it is clear that a significant number of the resulting fits will 
be very poor.  In defining ``plausible'' fit results, we must quantify 
what humans have been intuitively doing across three decades of radial velocity 
fitting.  We therefore establish five criteria that must be passed for a 
given eccentric-planet fit to be deemed plausible.

\begin{itemize}
\item The fitted eccentricity must be at least $3\sigma$ from zero. 
\medskip
\item The fitted velocity amplitude $K$ must be at least four times 
its own uncertainty: $K/\sigma_K > 4.0$.  This is derived from the NASA 
Exoplanet Archive, in which 95\% of confirmed radial-velocity-detected 
planets satisfy this criterion.
\medskip
\item The fitted velocity amplitude must be at least 1.23
times larger than the rms scatter about the fit: $K/\mathrm{rms} > 
1.23$.  As above, we choose this limit as that which holds true for 95\% 
of NASA Exoplanet Archive confirmed planets.
\medskip
\item The fitted period must be less than 1.5 times the total duration 
of the observations.  This is derived from noting that virtually no 
radial velocity planet discoveries are published with less than about 
0.7 orbital cycles of observations (cf. Figure 4 of \citet{jupiters}).
\medskip
\item The rms of the fit must be less than three times the mean measurement uncertainty: $\mathrm{rms}/\bar{\sigma} < 3.0$.

\end{itemize}

Whilst the final criterion is admittedly somewhat arbitrary, it eliminates 
obviously bad fits characterised by large residual scatter.  Given that 
the input data always contain two signals, and we fit for only one, it 
is reasonable to expect a large number of instances in which \texttt{RVLIN} 
chooses one periodicity and arrives at a ``best fit'' with an 
abominably large scatter.

Table~\ref{tab:fooled} briefly summarises the plausible 
single-eccentric fits resulting from the 12 trial scenarios described 
above, after applying these criteria. Scenario I, the double-circular 
configuration, resulted in 13.08\% plausible single-eccentric fits 
for the 2:1 period ratio ($P_2=$200 days), but none for the 2.17:1 period ratio.  Increasing the realism by drawing timestamps from real observations produced 19.04\% 
and 4.94\% such plausible fits for the 2:1 and 2.17:1 period ratios, respectively.  Similar results were achieved in Scenario II, using the ``slightly eccentric'' $e=0.1$ configuration.

Scenario III, where the outer period $P_2$ was drawn from the set of 
real planets, yielded the largest number of plausible single-eccentric 
fits, with 15.27\% (IIIa), 2.89\% (IIIb), 20.49\% (IIIc), and 9.30\% 
(IIId).  The detailed characteristics of these plausible fits are 
discussed further in the next subsection.

\begin{table}
        \centering
        \caption{``Plausible'' single-eccentric fit results from 10,000 trials}
        \label{tab:fooled}
        \begin{tabular}{llc}
                \hline
                Scenario & Number & Mean eccentricity \\
                \hline
                Ia & 1308 & 0.31$\pm$0.06 \\
                Ib & 0 & -- \\
                Ic & 1904 & 0.35$\pm$0.11 \\
                Id & 494 & 0.32$\pm$0.15 \\
                IIa & 1238 & 0.31$\pm$0.07 \\
                IIb & 0 & -- \\
                IIc & 2004 & 0.35$\pm$0.11 \\
                IId & 548 & 0.33$\pm$0.14 \\
                IIIa & 1527 & 0.31$\pm$0.10 \\
                IIIb & 289 & 0.28$\pm$0.16 \\
                IIIc & 2049 & 0.31$\pm$0.13 \\
                IIId & 930 & 0.29$\pm$0.16 \\
                \hline
        \end{tabular}
\end{table}


\subsection{Plausible Eccentric Single Solutions}

\begin{figure*}
	\includegraphics[width=\linewidth]{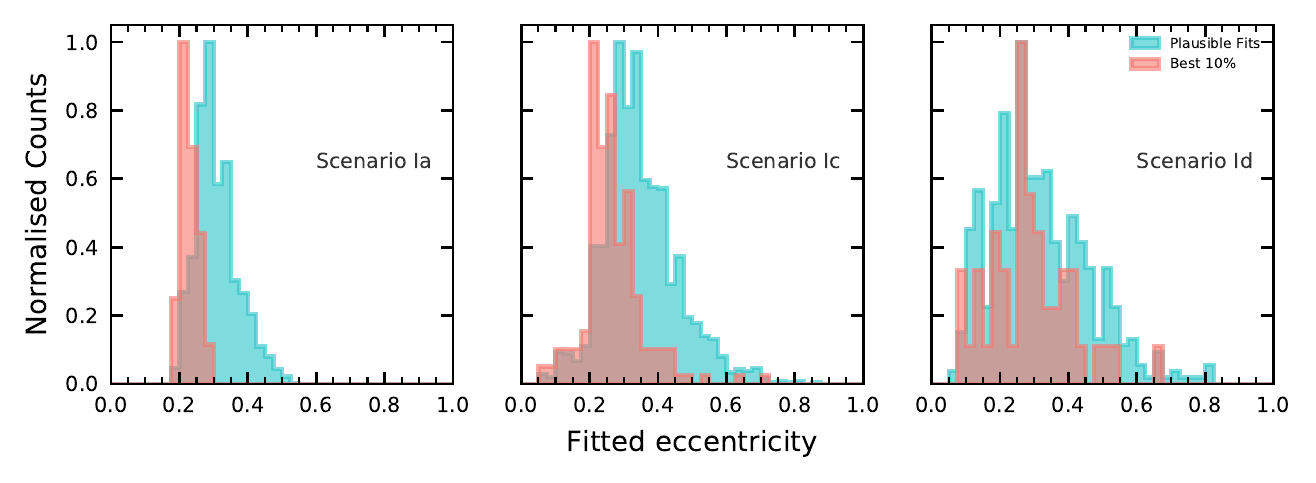}
    \caption{Distribution of fitted eccentricities for the ``plausible'' 
single-eccentric fits found in Scenario Ia, Ic, and Id.}
    \label{fig:eccdist1}
\end{figure*}

\begin{figure*}
	\includegraphics[width=\linewidth]{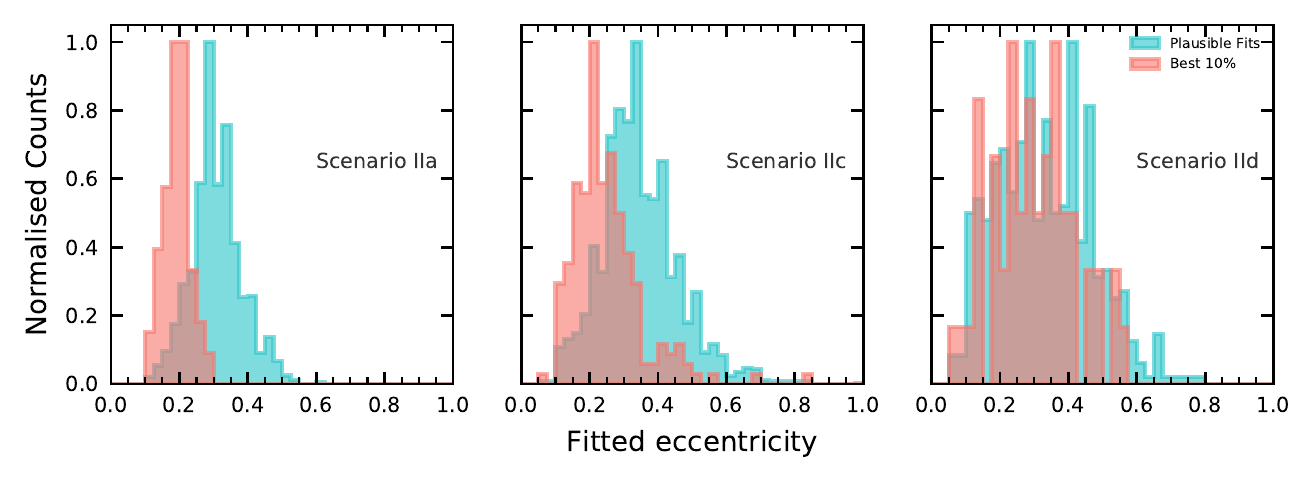}
    \caption{Same as Figure~\ref{fig:eccdist1}, but for Scenario IIa, 
IIc, and IId.}
    \label{fig:eccdist2}
\end{figure*}

\begin{figure*}
	\includegraphics[width=\linewidth]{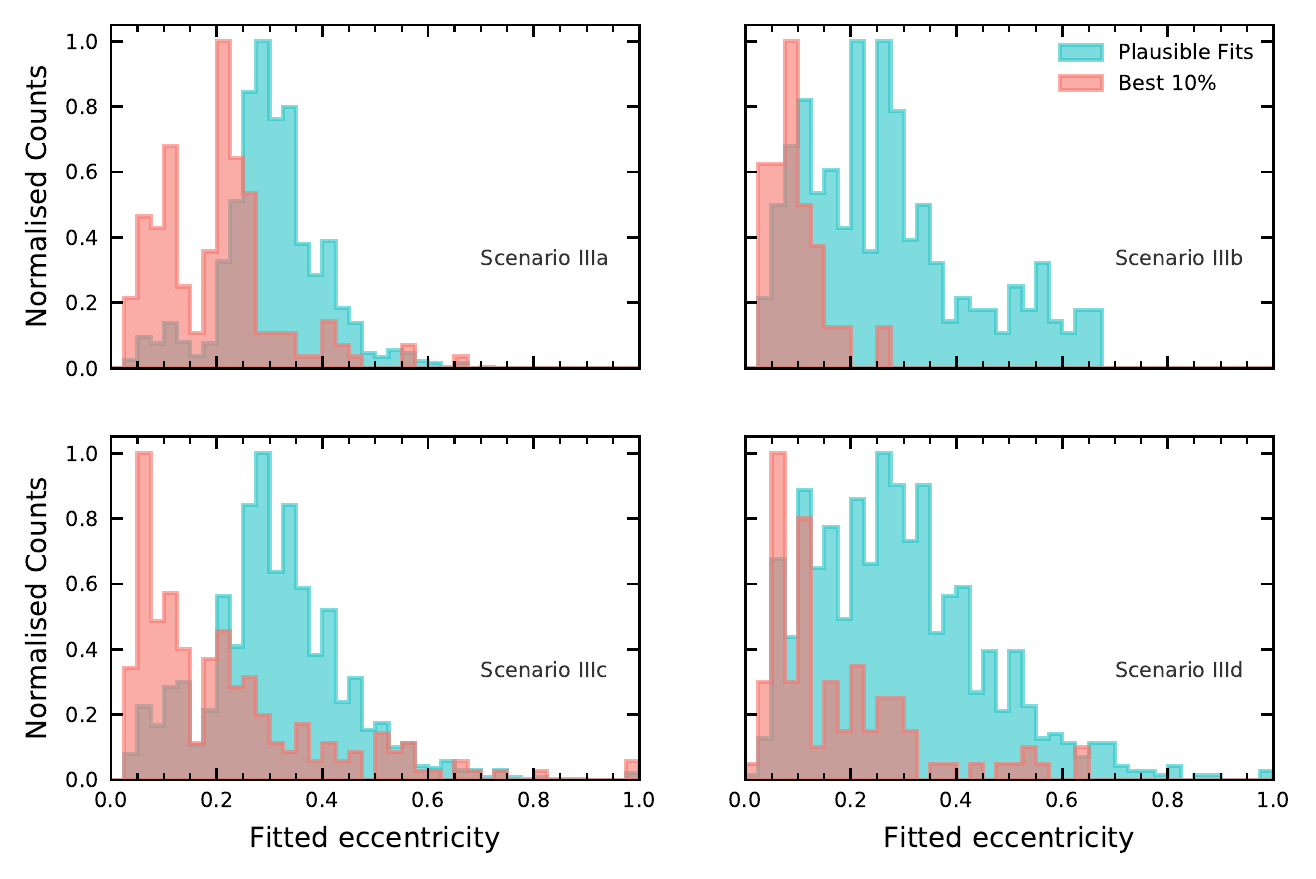}
    \caption{Same as Figure~\ref{fig:eccdist1}, but for Scenario IIIa, 
IIIb, IIIc, and IIId.}
    \label{fig:eccdist3}
\end{figure*}

In this subsection, we investigate the properties of the plausible single-eccentric fits in more detail.  Figures~\ref{fig:eccdist1}--\ref{fig:eccdist3} show as green histograms the distribution of fitted eccentricities obtained by the \texttt{RVLIN} single-planet fits which passed all five criteria described above.  Overplotted as red histograms are the ``best'' 10\% of plausible fits, those with the smallest ratio of rms scatter to mean measurement uncertainty.  This selection resulted in fits with rms values of 2.9-5.1\,\ms\ (where the mean measurement uncertainty is $\sim$3.2\,\ms).  These are fits which are most likely to convince the observer that a single eccentric planet is a good fit to the data, when in reality two planetary signals are ensconced within.

In Scenario I (periods fixed at 200:100 or 217:100 days), the bulk of the best fits were generally restricted to eccentricities between 0.2 -- 0.4, with a handful of examples out to $e\sim 0.6$ for Ic and Id, which sampled the simulated velocities using time series from real stars.  Notably, the 2.17:1 arrangement (Scenario Ib; no histogram possible) never produced a reasonable single-eccentric solution.  Scenario II, in which slightly eccentric ($e=0.1$) pairs of input planets were tested, gave very similar results to Scenario I.  Scenario III was the most realistic, with periods drawn from real radial velocity planets, and resulted in the highest fraction of plausible single-eccentric fits.  As in the previous trials, the best fits clustered at lower eccentricities ($e<0.2$) but developed a long tail extending even to $e>0.9$ for Scenarios IIIc and IIId (Figure~\ref{fig:eccdist3}).  Closer inspection revealed that the extreme examples were characterised by pathologies such as (1) reaching the maximum allowed period (10,000 days) with an error bar of more than 200\%, or (2) amplitudes $K>500$\,\ms\ with large uncertainties, likely driven by phase gaps in the time series (as the observation times were taken from real data).  These ``best'' fits were selected strictly by the lowest rms scatter, with no human intervention as yet, and so such oddities are to be expected.  We explore the cause of these peculiarities further in the next section.

\begin{figure}
\includegraphics[width=\columnwidth]{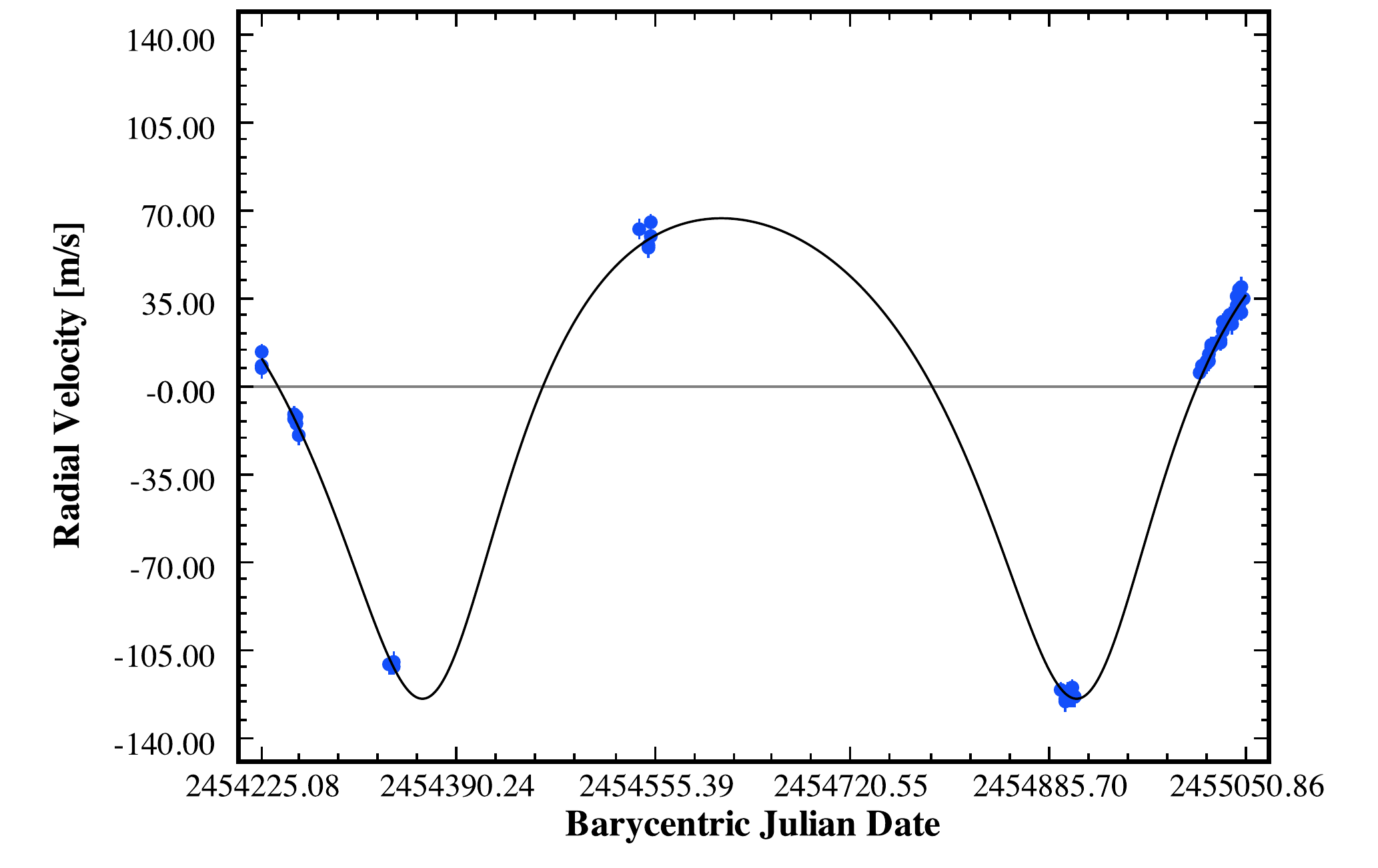}
\caption{Example of a simulated data set for which the circular-double combination can be plausibly fit with a single eccentric model.  Shown is an example from Scenario IIId, chosen from the ``best'' 10\% subset, to illustrate the pathology we investigate here.  This fit has a period of 548.8 days, $e=0.31\pm$0.03, and an rms of 2.56\,\ms.  The true injected data are two circular orbits with periods of 580.7 and 267.6 days. }
\label{fig:dataplot}
\end{figure}

Figure~\ref{fig:dataplot} shows an example of a simulated data set that yielded a ``good'' single-eccentric fit, with $e=0.31\pm0.03$, $K=95.6\pm0.9$\,\ms, and an rms of 2.56\,\ms.

\section{Discussion and Conclusions}
\label{Conclusions}

In this work, we consider the problematical false positive single-planet solutions that can arise from poorly sampled radial velocity observations of systems containing two exoplanets moving on near-circular orbits.  With poor sampling and noisy data, the analysis of such systems can often return a convincing single planet solution, with that planet moving on an orbit with moderate eccentricity. It is likely that a number of such near-circular exoplanetary couplets remain undiscovered among the many 'confirmed' single, moderately eccentric exoplanets.

The results of our analysis demonstrate the range of eccentricities for which it is conceivable that a circular two-planet system can masquerade as an eccentric single planet.  Particularly for the most realistic trials (IIIc and IIId), there was no obvious cutoff eccentricity beyond which circular-double systems failed to produce plausible single-eccentric fits.  However, in reality, it seems reasonable to expect such a threshold to exist.  This is due to the shape of the radial velocity orbit departing so far from sinusoidal that it realistically cannot be reproduced by only two circular components, which is the problem at the heart of this degeneracy (as noted in Equation 1 of \citealt{boisvert18}).  We see evidence of such a limit for the 'a' and 'b' subsets, in which the sampling turned out to be less realistic than for the `c' and `d' scenarios.  


The grand mean fitted eccentricity across all scenarios (Table~\ref{tab:fooled}) is $e=0.32\pm$0.12, with a 95\% confidence interval of [0.09, 0.59].  We take this to be the ``sweet spot'' (or danger zone) for deceptive configurations.  If we consider only the ``best'' 10\% of the plausible single-eccentric fits, we derive a grand mean $e=0.23\pm$0.12, with a 95\% confidence interval of [0.05, 0.53].  Our results therefore show that more eccentric planet candidates ($e\gtsimeq$0.5) are exceedingly unlikely to be mimics due to this degeneracy.  

\begin{figure*}
	\includegraphics[width=\linewidth]{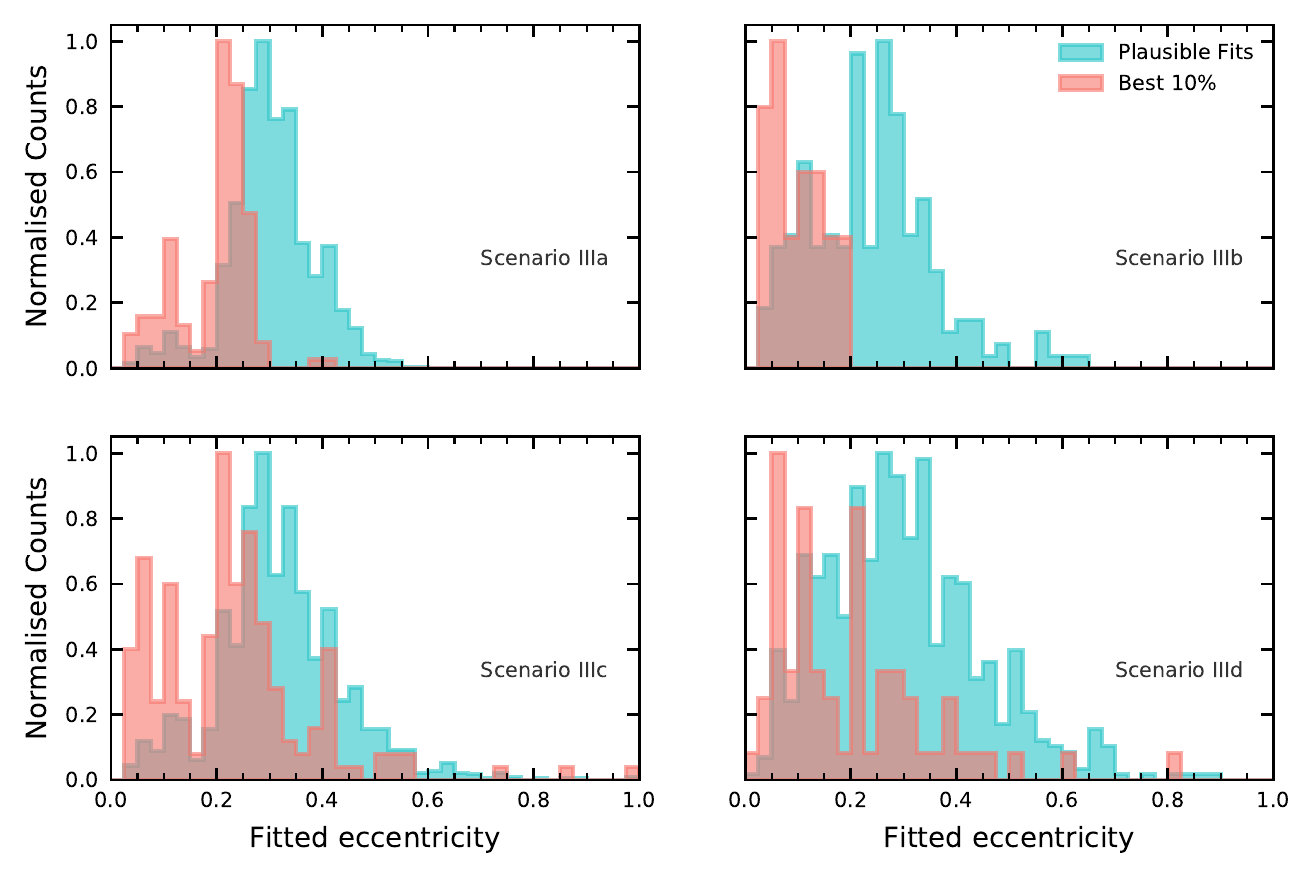}
    \caption{Same as Figure~\ref{fig:eccdist3}, but with the failed data sets excluded.}
    \label{fig:eccdist4}
\end{figure*}

As noted above, the best such fits (in an rms sense) sometimes exhibited pathologies such as extreme amplitudes brought on by phase gaps, or fitted periods at the upper boundary and with outsized error bars.  In Scenario III, up to 9\% of trials failed even to recover the two injected planets when subjected to a proper double-Keplerian fit (Table~\ref{tab:sanitycheck}).  We repeated the five tests described in Section 3 for the subset of trials which passed the sanity checking (see Section 2.4).  We show the results for this new set of ``plausible'' results in Figure~\ref{fig:eccdist4}.  For Scenarios IIIc and IIId, which were most affected, we indeed see that all but 2-3 of the $e>0.6$ fits are eliminated.  A handful remain, but the vast majority of ``good'' single-eccentric fits have $e\ltsimeq$0.4.  This is again consistent with our result that the ``danger zone'' lies generally in the range between $e\sim$0.2 and $e\sim$0.4.

The results presented herein are also relevant to exoplanet searches that utilise the transit method. The relative lack of orbital phase coverage from transiting planets without accompanying RV data makes the reliable extraction of orbital eccentricity from transit light curves a challenging endeavour \citep[e.g.][]{kipping08,vaneylen15}.  Statistical studies of orbital eccentricities derived from {\it Kepler} discoveries have found the eccentricity distribution of transiting planets to be consistent with that from the RV exoplanet population \citep{kane12}.  It was further concluded by \citet{kane12} and later by \citet{vaneylen15} that there is a negative correlation of orbital eccentricity with planet size, particularly for those planets in compact systems. These factors underline both the need for complementary RV observations of transiting planets to reliably extract orbital eccentricities, and the potential degeneracy of circular orbits in the terrestrial regime.  Exoplanet discoveries from the Transiting Exoplanet Survey Satellite ({\it TESS}) are expected to primarily be in relation to relatively bright host stars where such a complement of precision photometry and RV data will be far more accessible than for the {\it Kepler} systems \citep{ric15}.  In particular, extended mission scenarios for the {\it TESS} mission, such as those described by \citet{sul15}, allow for the detection of longer-period planets that will be more likely to have larger eccentricities.

Through the course of this work, we have demonstrated the impact of poor sampling on the veracity of convincing radial velocity exoplanet detections.  Our work reveals the regime for which most caution should be exercised when considering whether a single planet fit to observational data is a reflection of the true reality of the system in question.  At the same time, however, our results add credibility to the detections of exoplanets moving on highly eccentric orbits, such as HD\,80606b \citep[$e = 0.933 \pm 0.001$;][]{ecc1}, HD\,4113b \citep[$e = 0.903 \pm 0.005$;][]{ecc2} and HD\,76920b \citep[$e = 0.856 \pm 0.009$;][]{76920}.  Whilst such planets are clearly oddities, our results suggest that they are not false-positive, ghost planets.  Indeed, our results suggest that deceptive planetary couplets will rarely, if ever, masquerade as single planets with orbital eccentricities greater than $e\sim\,0.6$, consistent with the results of \citet{kurster15} as shown in their Figure 4.

In future work, we intend to extend this analysis still further, examining a wider range of planetary couplet period ratios, orbital eccentricities, and masses.  In addition, it is interesting to consider whether similar effects would result from deceptive triplets or quadruplets - in other words, whether it might be possible to mimic a signal of arbitrarily large eccentricity through the superposition of a given number (N > 2) of signals resulting from planets moving on near-circular orbits.


\section*{Acknowledgements}

CB is supported by Australian Research Council Discovery Grant DP170103491. JC is supported by an Australian Government Research Training Program (RTP) Scholarship. This research has made use of NASA's Astrophysics Data System (ADS), and the SIMBAD database, operated at CDS, Strasbourg, 
France. This research has made use of the NASA Exoplanet Archive, which is operated by the California Institute of Technology, under contract with the National Aeronautics and Space Administration under the Exoplanet Exploration Program.






\bsp	
\label{lastpage}
\end{document}